%% file: twolooptrispectrum.tex
\documentclass[a4paper,11pt]{article}
\usepackage{pos}
\usepackage{physics}
\usepackage{float}
\usepackage{subcaption}
\usepackage{comment}
\usepackage{tikz}
\usepackage{circuitikz}
\usetikzlibrary{shapes.misc, external}
\usetikzlibrary{arrows.meta,decorations.markings}
\title{Efficient Numerical Evaluation of a Two-Loop Contribution to the Dark-Matter Trispectrum}
\ShortTitle{Efficient Numerical Evaluation of a Two-Loop Trispectrum Contribution}

\author*[a]{Andrea Favorito}

\affiliation[a]{Institute for Theoretical Physics, ETH Zurich, 8093 Zurich, Switzerland}

\emailAdd{afavorito@phys.ethz.ch}

\abstract{
We study a two-loop contribution to the dark-matter trispectrum and evaluate it numerically using an infrared-safe integrand. The calculation is organized as an expansion around a fixed reference cosmology: the linear matter power spectrum of the target cosmology is written as a rescaled reference spectrum plus a small difference, and the trispectrum is expanded perturbatively in this difference. For the external momentum configuration considered here, truncating the expansion at third order reproduces the full numerical result with sub-percent accuracy over the range of scales studied, while higher-order terms are strongly suppressed. This reorganization reduces the number of cosmology-independent building blocks that must be computed compared with direct basis decompositions of the linear power spectrum. This provides a practical route to faster evaluations of higher-loop and higher-multiplicity correlators in the Effective Field Theory of Large-Scale Structure.
}

\FullConference{17th International Symposium on Radiative Corrections: Applications of Quantum Field Theory to Phenomenology (RADCOR2025)\\
5-10 October 2025\\
Puri, India\\}

\renewcommand{\logo}{\relax}

\newcommand{\Plin}{\mathcal{P}_{\rm lin.}}
\newcommand{\DPlin}{\Delta \Plin}
\newcommand{\ie}{\emph{i.e.}}

\newcommand{\Ncal}{{\cal N}}
\newcommand{\Tcal}{{\cal T}}
\newcommand{\Dcal}{{\cal D}}
\newcommand{\vk}{\vec k}
\newcommand{\vq}{\vec q}

\newcommand{\vQ}{\vec Q}
\newcommand{\diag}{T_{2233,\rm(VI)}}
\newcommand{\diagSup}[1]{T^{#1}_{2233, \rm (VI)}}
\newcommand{\loopmeasure}[1]{\frac{\dd^3 #1}{(2 \pi)^3}}

\begin{document}
\maketitle

\section{Introduction}
\label{sec:intro}

The statistical precision of large-scale structure measurements is improving rapidly, pushing analyses beyond the regime where tree-level or one-loop predictions are sufficient
\cite{BOSS:2016wmc, DESI:2024jxi,DAmico:2022osl}.
In the Effective Field Theory of Large-Scale Structure (EFTofLSS) \cite{Baumann:2010tm, Carrasco:2012cv}, fully exploiting the constraining power of these measurements requires repeated evaluations of loop corrections across cosmological parameter space in likelihood analyses. At multi-loop order and for higher-point correlators, these corrections become complicated multi-dimensional momentum integrals, making direct numerical evaluation at each likelihood call computationally unfeasible.

A common approach to accelerate inference is to approximate the linear power spectrum using a finite expansion in $N$ basis functions and to precompute the corresponding cosmology-independent loop building blocks. This idea underlies FFTLog-based methods~\cite{Simonovic:2017mhp}, optimized decompositions such as COBRA~\cite{Bakx:2024zgu, Bakx:2025pop, Bakx:2025jwa}, and mappings of EFTofLSS loop integrals to QFT-like topologies~\cite{Anastasiou:2022udy,DAmico:2022osl}. The bottleneck is basis size: for a diagram with $M$ internal power spectrum insertions, the number of precomputed integrals scales as $\mathcal{O}(N^M)$, which becomes prohibitive at higher loop order and multiplicity.

This motivates strategies that reduce the number and complexity of the required cosmology-independent integrals at fixed accuracy. One such strategy, introduced in Ref.~\cite{Anastasiou:2025jsy}, evaluates loop corrections numerically while expanding the cosmology dependence around a fixed reference model. This reorganization requires only a smaller set of cosmology-independent building blocks, enabling efficient extensions to higher-loop and higher-multiplicity correlators. Below we summarize the method and apply it to a two-loop contribution to the dark-matter trispectrum, the connected four-point correlation function in momentum space.

\section{Setup}
\label{sec:setup}

Let $\Plin^{[j]}(k)$ denote the linear matter power spectrum for the set of cosmological parameters labeled by $j$, evaluated at fixed redshift%
\footnote{In practice, the inputs $\Plin^{[j]}$ are obtained by solving the linearized Einstein--Boltzmann system with standard public codes such as CLASS~\cite{Blas:2011rf}.}%
\textsuperscript{\!,}%
\footnote{Throughout, we evaluate the correlator at redshift $z=0$ and adopt the Einstein--de Sitter (EdS) approximation for the time dependence of the perturbation theory kernels. With this choice, we suppress explicit growth-factor dependence (e.g.\ $D^n$ factors) and evaluate all kernels consistently at the same time.}. For a target cosmology and a fixed reference cosmology (labelled ``$0$''), we write
\begin{equation}
\label{eq:DeltaP}
    \Plin^{[j]}(k)
    =
    \Ncal^{[j]} \Plin^{[0]}(k) + \DPlin^{[j]}(k)
    \qquad \text{where} \qquad
    \Ncal^{[j]} = \frac{\max_k \Plin^{[j]}(k)}{\max_k \Plin^{[0]}(k)} \ ,
\end{equation}
and treat the correlator as an expansion in $\DPlin^{[j]}(k)$, exploiting that $\DPlin^{[j]}$ is small for cosmologies close to the reference model.
Compared to related two-loop EFTofLSS calculations~\cite{Bakx:2025jwa}, this strategy reduces the complexity of the required cosmology-independent loop building blocks, lowering the maximal tensor rank that must be computed.

\begin{figure}[ht]
\centering
\input{tikz}
\caption{Two-loop ``sunset-like'' topology contributing to the $2233$ contraction of the matter trispectrum.}
\label{fig:topo}
\end{figure}
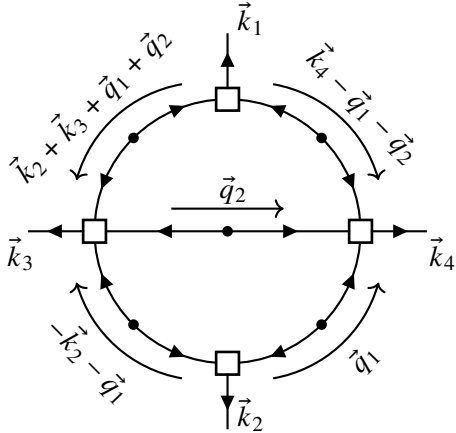

In this work we apply this method to a non-trivial two-loop topology entering the dark-matter four-point function (trispectrum), reminiscent of the ``sunset'' graph in perturbative Quantum Field Theories (pQFTs). The contribution we consider arises from the component with two overdensity fields expanded to second order and two fields expanded to third order, \ie, $\langle \delta^{(2)}\delta^{(2)}\delta^{(3)}\delta^{(3)}\rangle$, which motivates the label ``$2233$'' used below. 
The topology is shown in Fig.~\ref{fig:topo}. It contains five internal lines and four vertices, which implies that its integrand involves five factors of $\Plin$ and four symmetrized dark matter kernels $F_n^{(s)}$ (see, e.g., \cite[Sec.~2]{Anastasiou:2025jsy}). For external momenta (wavenumbers) $\vk_i$ satisfying momentum conservation $\vk_4=-\vk_1-\vk_2-\vk_3$, we define
\begin{equation}
    \label{eq:integral}
    \diag(\vk_1,\vk_2,\vk_3,\vk_4)
    \;=\;
    \int \loopmeasure{q_1}\int \loopmeasure{q_2}\;
    \Tcal_{2233,{\rm (VI)}}(\vq_1,\vq_2,\vk_1,\vk_2,\vk_3,\vk_4)\, .
\end{equation}
The integrand is obtained by symmetrizing over the six $(2,2)$-shuffles,
\begin{equation}
\label{eq:integrand_full}
\begin{aligned}
    \Tcal_{2233,{\rm (VI)}}(\vq_1,\vq_2,\vk_1,\vk_2,\vk_3,\vk_4)
    \;=\;
    \sum_{\sigma\in{\rm Sh}(2,2)}
    \Tcal^{\rm unsym.}_{2233,{\rm (VI)}}\!\left(\vq_1,\vq_2,\vk_{\sigma(1)},\vk_{\sigma(2)},\vk_{\sigma(3)},\vk_{\sigma(4)}\right)\, ,
\end{aligned}
\end{equation}
where ${\rm Sh}(2,2)$ denotes the set of $(2,2)$-shuffles in the shuffle algebra, explicitly
$$
{\rm Sh}(2,2)=\{[1,2,3,4],\,[1,3,2,4],\,[1,4,2,3],\,[2,3,1,4],\,[2,4,1,3],\,[3,4,1,2]\}\, .
$$
The unsymmetrised contribution reads
\begin{equation}
\label{eq:integrand_ref}
\begin{aligned}
    \Tcal^{\rm unsym.}_{2233, \rm (VI)} (\vq_1, \vq_2, \vk_1, \vk_2, \vk_3, \vk_4)
    =&
    144\;
    \Plin(\vQ_1)
    \Plin(\vQ_2)
    \Plin(\vQ_3)
    \Plin(\vq_1)
    \Plin(\vq_2)
    \\
    &\times
    F_3^{(s)} (\vQ_1, - \vq_2, \vQ_3)
    F_3^{(s)} (\vQ_2, \vq_1, \vq_2)
    \\
    &\times
    F_2^{(s)} (- \vQ_1, -\vQ_2)
    F_2^{(s)} (-\vQ_3, - \vq_1)\, ,
\end{aligned}
\end{equation}
with momentum combinations
\begin{equation} \label{eq:Qs}
    \vQ_1=\vk_2+\vk_3+\vq_1+\vq_2,\qquad
    \vQ_2=\vk_4-\vq_1-\vq_2,\qquad
    \vQ_3=-\vk_2-\vq_1\, .    
\end{equation}

If one attempted to map this integral to a standard pQFT form following Ref.~\cite{Anastasiou:2022udy}, one would encounter a two-loop topology with up to seven propagators: five massive lines with non-degenerate masses and two massless lines, making an analytic treatment prohibitively complicated. In contrast, the present approach allows a straightforward numerical evaluation. While we do not construct a full basis of cosmology-independent integrals here, we instead compute the diagram for a single reference cosmology, with linear power spectrum $\Plin^{[0]}$, and study the convergence of the $\DPlin$ expansion around it. As we will show, the accuracy targets of interest can be reached with a lower maximal rank of the tensor integrals than what is typically required by analytic approaches~\cite{Anastasiou:2022udy,Simonovic:2017mhp} or alternative numerical strategies~\cite{Bakx:2025jwa}.

For a stable numerical evaluation, it is essential to work with an integrand representation that is manifestly finite in all infrared (IR) and ultraviolet (UV) regions. According to the UV power counting defined in Ref.~\cite[Sec.~4]{Anastasiou:2025jsy}, the integrand in Eq.~\eqref{eq:integrand_ref} is convergent in each of the ultraviolet limits. The infrared structure is more subtle: while the full \emph{integral} is IR-finite, the \emph{integrand} contains an entangled set of local integrable IR singularities that can severely degrade numerical performance. To disentangle these regions and construct an IR-safe integrand suitable for numerical integration, we apply the same rearrangement and removal procedure as in Ref.~\cite[Sec.~5]{Anastasiou:2025jsy}. We review this construction in the next section.

\section{Infrared safe integrand}

We use the same power-counting rules as Ref.~\cite[Sec.~5]{Anastasiou:2025jsy}.
Specifically, to probe a soft region associated with an internal momentum $q$, we rescale
$q \mapsto \delta q$, and study the behavior of the integrand as $\delta \to 0^+$. 
If the integrand together with the integration measure scales as $\delta^{\alpha}$ in this limit, the exponent $\alpha$ 
provides a convenient measure of the degree of (non-)integrability of the corresponding soft configuration.
In what follows we assume the infrared scaling of the linear power spectrum,
\begin{equation}
    \Plin \left( q \delta \right) = C_{\rm IR} q \delta + \mathcal{O} \left( \delta^2 \right),
\end{equation}
where $C_{\rm IR}$ is a cosmology-dependent constant\footnote{More generally, in the infrared one has $\Plin(k) \propto k^{n_s}$ \cite{Dodelson:2003ft}, with $n_s \approx 0.965$ \cite{Planck:2018jri}. Setting $n_s=1$ is a simplifying assumption; however, the argument below applies unchanged for any $n_s>-1$.}.

A direct expansion of the kernels in the soft limits shows that the integrand in Eq.~\eqref{eq:integrand_ref} develops local IR singularities. Although these singularities are integrable once combined with the integration measure, they still appear as pointwise poles in the integrand and can therefore degrade the stability of the numerical integration. In particular, the integrand contains ``single'' singularities when any one of the internal momenta becomes soft, and ``double'' singularities when any pair of internal momenta becomes soft simultaneously. After including the integration measure, the soft momenta acquire additional powers, so some of these soft configurations become finite and therefore do not require any further subtraction. With the momentum routing chosen in Fig.~\ref{fig:topo}, the remaining singular regions are associated with the combinations $\vQ_1$, $\vQ_2$, and $\vQ_3$ defined in Eq.~\eqref{eq:Qs}. A convenient way to cancel the corresponding IR behavior at the integrand level is to multiply by an IR measure that vanishes in all of these soft limits,
\begin{equation}
    \mu(\vQ_1,\vQ_2,\vQ_3) = \vQ_1^2\,\vQ_2^2\,\vQ_3^2 \ .
\end{equation}

Multiplying the integrand in Eq.~\eqref{eq:integrand_ref} by the IR measure $\mu$ would, by itself, change the value of the integral in Eq.~\eqref{eq:integral}. Following Ref.~\cite[Sec.~5]{Anastasiou:2025jsy}, we instead introduce $\mu$ through a partition of unity. Define the normalization
\begin{equation}
\label{eq:normalisation}
\begin{aligned}
    \Dcal \;=&\;
    \mu(\vQ_1, \vQ_2, \vQ_3)
    + \mu(\vQ_1, \vQ_2, \vq_1)
    + \mu(\vQ_1, \vQ_3, \vq_1)
    + \mu(\vQ_1, \vQ_3, \vq_2)
    \\
    &\qquad
    + \mu(\vQ_1, \vq_1, \vq_2)
    + \mu(\vQ_2, \vQ_3, \vq_1)
    + \mu(\vQ_2, \vQ_3, \vq_2)
    + \mu(\vQ_2, \vq_1, \vq_2)\, ,
\end{aligned}
\end{equation}
so that the identity
\begin{equation}
\label{eq:partition_unity}
\begin{aligned}
    1
    \;=\;
    \frac{1}{\Dcal}\Big(
    &\mu(\vQ_1, \vQ_2, \vQ_3)
    + \mu(\vQ_1, \vQ_2, \vq_1)
    + \mu(\vQ_1, \vQ_3, \vq_1)
    + \mu(\vQ_1, \vQ_3, \vq_2)
    \\
    &\qquad
    + \mu(\vQ_1, \vq_1, \vq_2)
    + \mu(\vQ_2, \vQ_3, \vq_1)
    + \mu(\vQ_2, \vQ_3, \vq_2)
    + \mu(\vQ_2, \vq_1, \vq_2)
    \Big)
\end{aligned}
\end{equation}
holds trivially by construction.

Inserting Eq.~\eqref{eq:partition_unity} into Eq.~\eqref{eq:integral} and performing the appropriate momentum shifts, we obtain an equivalent representation of the reference integrand as a sum of eight terms,
\begin{equation}
    \label{eq:ir-safe-integrand_terms}
    \begin{aligned}
        \Tcal^{\rm unsym., (1)}_{2233, \rm (VI)}(\vq_1, \vq_2, \vk_1, \vk_2, \vk_3, \vk_4) &= \frac{1}{\Dcal_1} \mu(\vQ_1, \vQ_2, \vQ_3) \, \Tcal^{\rm unsym.}_{2233, \rm (VI)} (\vq_1, \vq_2, \vk_1, \vk_2, \vk_3, \vk_4) \ , \\ 
        \Tcal^{\rm unsym., (2)}_{2233, \rm (VI)}(\vq_1, \vq_2, \vk_1, \vk_2, \vk_3, \vk_4) &= \frac{1}{\Dcal_2} \mu(\vQ_4, \vQ_8, \vQ_3) \, \Tcal^{\rm unsym.}_{2233, \rm (VI)} (\vQ_3, \vq_2, \vk_1, \vk_2, \vk_3, \vk_4) \ , \\ 
        \Tcal^{\rm unsym., (3)}_{2233, \rm (VI)}(\vq_1, \vq_2, \vk_1, \vk_2, \vk_3, \vk_4) &= \frac{1}{\Dcal_3} \mu(\vQ_5, \vQ_9, \vQ_2)  \, \Tcal^{\rm unsym.}_{2233, \rm (VI)} (\vQ_2, \vq_2, \vk_1, \vk_2, \vk_3, \vk_4) \ , \\ 
        \Tcal^{\rm unsym., (4)}_{2233, \rm (VI)}(\vq_1, \vq_2, \vk_1, \vk_2, \vk_3, \vk_4) &= \frac{1}{\Dcal_4} \mu(\vQ_5, \vQ_6, \vQ_2)  \, \Tcal^{\rm unsym.}_{2233, \rm (VI)} (\vq_2, \vQ_2, \vk_1, \vk_2, \vk_3, \vk_4) \ , \\ 
        \Tcal^{\rm unsym., (5)}_{2233, \rm (VI)}(\vq_1, \vq_2, \vk_1, \vk_2, \vk_3, \vk_4) &= \frac{1}{\Dcal_5} \mu(\vQ_5, \vQ_6, \vQ_{10})  \, \Tcal^{\rm unsym.}_{2233, \rm (VI)} (\vQ_6, \vQ_{10}, \vk_1, \vk_2, \vk_3, \vk_4) \ , \\ 
        \Tcal^{\rm unsym., (6)}_{2233, \rm (VI)}(\vq_1, \vq_2, \vk_1, \vk_2, \vk_3, \vk_4) &= \frac{1}{\Dcal_6} \mu(\vQ_5, \vQ_4, \vQ_7)  \, \Tcal^{\rm unsym.}_{2233, \rm (VI)} (\vQ_7, \vq_2, \vk_1, \vk_2, \vk_3, \vk_4) \ , \\ 
        \Tcal^{\rm unsym., (7)}_{2233, \rm (VI)}(\vq_1, \vq_2, \vk_1, \vk_2, \vk_3, \vk_4) &= \frac{1}{\Dcal_7} \mu(\vQ_5, \vQ_6, \vQ_7) \, \Tcal^{\rm unsym.}_{2233, \rm (VI)} (\vq_2, \vQ_7, \vk_1, \vk_2, \vk_3, \vk_4) \ , \\ 
        \Tcal^{\rm unsym., (8)}_{2233, \rm (VI)}(\vq_1, \vq_2, \vk_1, \vk_2, \vk_3, \vk_4) &= \frac{1}{\Dcal_8} \mu(\vQ_5, \vQ_6, \vQ_{11}) \, \Tcal^{\rm unsym.}_{2233, \rm (VI)} (\vQ_6, \vQ_{11}, \vk_1, \vk_2, \vk_3, \vk_4) \ .
    \end{aligned}
\end{equation}
Here the additional momentum combinations are
\begin{equation}
\label{eq:Q4toQ11}
\begin{aligned}
    \vQ_4 &= \vk_3 - \vq_1 + \vq_2 \, , &
    \vQ_5 &= - \vk_1 - \vq_1 \, , &
    \vQ_6 &= -\vk_2 - \vq_2 \, , & \\
    \vQ_7 &= -\vk_2 - \vk_3 + \vq_1 - \vq_2 \, , &
    \vQ_8 &= \vk_2 + \vk_4 + \vq_1 - \vq_2 \, , &
    \vQ_9 &= - \vk_2 - \vk_4 + \vq_1 + \vq_2 \, , \\
    \vQ_{10} &= \vk_2 + \vk_4 - \vq_1 + \vq_2 \, , &
    \vQ_{11} &= -\vk_3 + \vq_1 + \vq_2 \, .
\end{aligned}
\end{equation}
The normalization factor $1/\Dcal$ is affected by the momentum shifts applied in each term, leading to eight corresponding denominators $1/\Dcal_i$. Their explicit expressions are given in Appendix~\ref{app:formulas}.
    
An explicit check shows that each term in Eq.~\eqref{eq:ir-safe-integrand_terms} yields an infrared-finite integral. Since Eq.~\eqref{eq:partition_unity} is an identity, this rearrangement does not change the value of the diagram in Eq.~\eqref{eq:integral}. Thus, we have
\begin{equation}
    \label{eq:ir-safe-integrand}
    \begin{aligned}
    \sum_{\sigma \in \mathrm{Sh}(2,2)} \sum_{i = 1}^8 \int \loopmeasure{q_1} \int \loopmeasure{q_2} \Tcal^{\rm unsym., (i)}_{2233, \rm (VI)} (\vq_1, \vq_2, \vk_{\sigma(1)}, \vk_{\sigma(2)}, &\vk_{\sigma(3)}, \vk_{\sigma(4)}) \\
     & \; = \diag (\vk_1, \vk_2, \vk_3, \vk_4) \ .
    \end{aligned}
\end{equation}

This construction can be extended to other EFTofLSS integrands. We leave a systematic discussion of this extension to future work.

\begin{figure}
    \centering
    \includegraphics[]{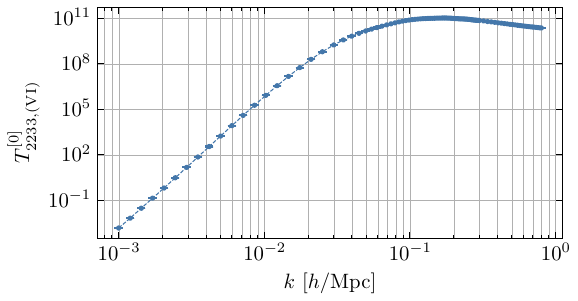}
    \caption{Numerical evaluation of $\diag(\vk_1,\vk_2,\vk_3,\vk_4)$ using the IR-safe integrand in Eq.~\eqref{eq:ir-safe-integrand}, for the reference (``Planck'') cosmology, with linear power spectrum $\Plin^{[0]}$, in the square configuration defined in Eq.~\eqref{eq:config}.}
    \label{fig:Planck_integral}
\end{figure}

With the IR-safe representation in Eqs.~\eqref{eq:ir-safe-integrand_terms}--\eqref{eq:ir-safe-integrand}, the two-loop integral in Eq.~\eqref{eq:integral} can be evaluated numerically using the \texttt{Vegas} routine~\cite{Lepage:1977sw,Lepage:2020tgj}, together with the extrapolation and mapping procedure described in Ref.~\cite[Sec.~6.1]{Anastasiou:2025jsy}. In Fig.~\ref{fig:Planck_integral} we show a representative result for the reference cosmology, taken to be the ``$0$'' model with cosmological parameters reported in Ref.~\cite[Sec.~4]{Anastasiou:2025jsy}. 
We consider the square configuration
\begin{equation}
    \label{eq:config}
    \vk_1 = (k, 0, 0) \, , \ \vk_2 = (0, k, 0)\, , \ \vk_3 = (-k, 0, 0)\, ,\ \vk_4 = (0, -k, 0)\ ,
\end{equation}
and vary $k$ over 99 values in the range $[10^{-3},\,0.8]$.

\section{Double perturbative expansion}

The integrals evaluated in the previous section correspond to the reference cosmology, labelled ``$0$'', introduced in Eq.~\eqref{eq:DeltaP}. We now extend the calculation to a target cosmology $j$ by inserting the expansion of Sec.~\ref{sec:setup}, \ie, by replacing each occurrence of $\Plin$ in Eq.~\eqref{eq:ir-safe-integrand} with Eq.~\eqref{eq:DeltaP}. Since the integrand contains five factors of $\Plin$, this produces $2^5=32$ terms for each shuffle of the external legs, which can be organized according to the number of $\DPlin$ insertions. In this way, for any cosmology $j$ we write
\begin{equation}
\label{eq:expanded}
\begin{aligned}
    \diagSup{[j]}(\vk_1,\cdots,\vk_4)
    \;=&\;
    \left(\Ncal^{[j]}\right)^5 \diagSup{[0]}(\vk_1,\cdots,\vk_4)
    + \left(\Ncal^{[j]}\right)^4 \Delta_1 \diagSup{[j]}(\vk_1,\cdots,\vk_4)
    \\
    &\;+\;
    \left(\Ncal^{[j]}\right)^3 \Delta_2 \diagSup{[j]}(\vk_1,\cdots,\vk_4)
    + \left(\Ncal^{[j]}\right)^2 \Delta_3 \diagSup{[j]}(\vk_1,\cdots,\vk_4)
    \\
    &\;+\;
    \Ncal^{[j]} \Delta_4 \diagSup{[j]}(\vk_1,\cdots,\vk_4)
    + \Delta_5 \diagSup{[j]}(\vk_1,\cdots,\vk_4)\, .
\end{aligned}
\end{equation}
Here, $\Delta_m \diagSup{[j]}$ (with $m=1,\dots,5$) denotes the sum of all contributions containing exactly $m$ insertions of $\DPlin^{[j]}$ and $5-m$ insertions of $\Plin^{[0]}$, including the corresponding momentum routings and the shuffle symmetrization. In particular, $\Delta_m \diagSup{[j]}$ is of order $(\DPlin)^m$.

Since $\DPlin^{[j]}$ is expected to be small for cosmologies close to the reference model, the higher-$m$ contributions should be increasingly suppressed. This motivates defining the truncated approximation to Eq.~\eqref{eq:expanded} at order $m$ as
\begin{equation}
\label{eq:expanded_truncated}
\begin{aligned}
    \diagSup{[j],(m)}(\vk_1,\cdots,\vk_4)
    \;\equiv&\;
    \left(\Ncal^{[j]}\right)^5 \diagSup{[0]}(\vk_1,\cdots,\vk_4) \\
    &\qquad + \sum_{i=1}^{m} \left(\Ncal^{[j]}\right)^{5-i}\, \Delta_i \diagSup{[j]}(\vk_1,\cdots,\vk_4)\, .
\end{aligned}
\end{equation}
This suggests that, for cosmologies sufficiently close to the reference model, truncation at $m=2$ or $m=3$ may already achieve sub-percent accuracy; we assess this expectation below.

\section{Conclusions}

\begin{figure}[H]
\centering
    \begin{subfigure}[b]{0.45\textwidth}
    \centering
    \includegraphics[]{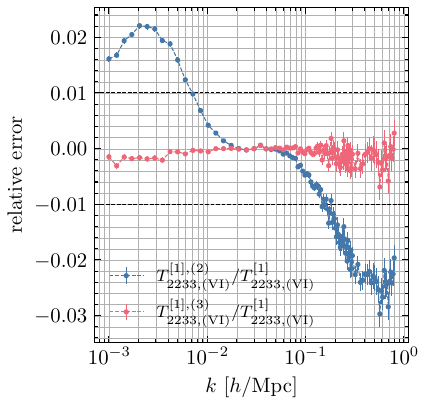}
    \caption{\label{fig:rel_err}}
    \end{subfigure}
\hfill
    \begin{subfigure}[b]{0.45\textwidth}
    \centering
    \includegraphics[]{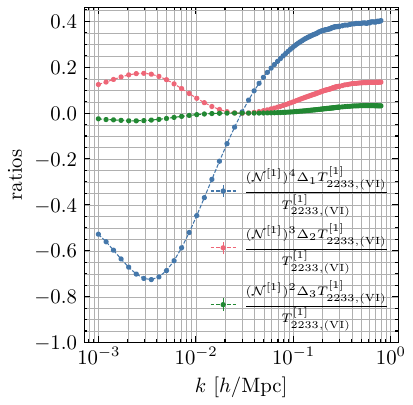}
    \caption{\label{fig:ratios}}
    \end{subfigure}
    \caption{(a) Relative error of the truncated expansions $\diagSup{[1],(2)}$ and $\diagSup{[1],(3)}$ with respect to the full numerical result, in the square configuration of Eq.~\eqref{eq:config}. (b) Ratios $\left(\Ncal^{[1]}\right)^{5-m}\Delta_m \diagSup{[1]}/\diagSup{[1]}$ illustrating the relative size of successive orders in the $\DPlin$ expansion.}
    \label{fig:comparison}
\end{figure}

To assess the convergence of the $\DPlin$ expansion, Fig.~\ref{fig:comparison} shows results for the benchmark cosmology $j=1$, with parameters given in Ref.~\cite[Sec.~7]{Anastasiou:2025jsy}. In the square configuration of Eq.~\eqref{eq:config} and over the full $k$ range considered, the expansion converges rapidly: the $m=3$ truncation agrees with the full numerical result at the sub-percent level, while the $m=2$ truncation shows percent-level deviations that reach a few percent at the upper end of the range (Fig.~\ref{fig:rel_err}).

To assess the relative importance of the successive corrections, Fig.~\ref{fig:ratios} shows the ratios
\begin{equation}
        \frac{\left(\Ncal^{[1]}\right)^{5-m}\Delta_m \diagSup{[1]}}{\diagSup{[1]}} \ .
\end{equation}
We find that the $m=3$ term gives at most a $\mathcal{O}(1\%)$--$\mathcal{O}(2\%)$ correction, consistent with the observed accuracy of the $m=3$ truncation and indicating that higher orders are strongly suppressed for the cosmologies of interest.

Beyond numerical convergence, this approach can substantially reduce the cost of cosmology emulation. In a direct basis decomposition of $\Plin$ for a diagram with five $\Plin$ insertions, the number of cosmology-independent building blocks scales as $\mathcal{O}(N^5)$, where $N$ is the number of basis functions. In the present organization, instead, the result is expanded in $\DPlin$, and one may choose a different fitting basis for $\DPlin$ at each order $m$. Since the higher-$m$ terms are progressively suppressed, they can be represented with less accurate, and therefore smaller, bases. If $N_m$ basis functions are used at order $m$, the cost of the corresponding contraction scales as $\mathcal{O}((N_m)^m)$.
In particular, retaining terms up to $m=3$ replaces the $\mathcal{O}(N^5)$ scaling of a direct fit by a leading higher-order cost of $\mathcal{O}(N_3^3)$, providing a clear route to reducing the number of required cosmology-independent building blocks for higher-loop and higher-multiplicity correlators.

\section*{Acknowledgments}
We would like to thank our collaborators on the project in Ref.~\cite{Anastasiou:2025jsy}, Charalampos Anastasiou, Matthew Lewandowski, Leonardo Senatore, and Henry Zheng. We acknowledge the use of the Euler cluster at ETH for much of the numerical computations. This work was supported by the Swiss National Science Foundation through its project
funding scheme (grant number 10001706). 

\appendix
\section{Normalization factors for the IR-safe decomposition}
\label{app:formulas}

For completeness, we report here the explicit expressions of the normalization factors $\Dcal_i$ appearing in Eq.~\eqref{eq:ir-safe-integrand_terms}. They are obtained from the partition of unity in Eq.~\eqref{eq:partition_unity} after applying the corresponding momentum shifts for each term $i=1,\ldots,8$.

\begin{equation}
\begin{aligned}
\Dcal_1 &= \Dcal\\
\Dcal_2 &= \mu(\vq_1,\vq_2,\vQ_4)+\mu(\vq_1,\vq_2,\vQ_8)+\mu(\vq_1,\vQ_3,\vQ_4)+\mu(\vq_1,\vQ_3,\vQ_8)+ \\ 
        & \qquad + \mu(\vq_1,\vQ_4,\vQ_8)+\mu(\vq_2,\vQ_3,\vQ_4)+\mu(\vq_2,\vQ_3,\vQ_8)+\mu(\vQ_3,\vQ_4,\vQ_8)\\
\Dcal_3 &= \mu(\vq_1,\vq_2,\vQ_2)+\mu(\vq_1,\vq_2,\vQ_9)+\mu(\vq_1,\vQ_2,\vQ_5)+\mu(\vq_1,\vQ_2,\vQ_9)+ \\ 
        & \qquad + \mu(\vq_1,\vQ_5,\vQ_9)+\mu(\vq_2,\vQ_2,\vQ_5)+\mu(\vq_2,\vQ_5,\vQ_9)+\mu(\vQ_2,\vQ_5,\vQ_9)\\
\Dcal_4 &= \mu(\vq_1,\vq_2,\vQ_2)+\mu(\vq_1,\vq_2,\vQ_5)+\mu(\vq_1,\vq_2,\vQ_6)+\mu(\vq_1,\vQ_2,\vQ_6)+ \\ 
        & \qquad + \mu(\vq_1,\vQ_5,\vQ_6)+\mu(\vq_2,\vQ_2,\vQ_5)+\mu(\vq_2,\vQ_5,\vQ_6)+\mu(\vQ_2,\vQ_5,\vQ_6)\\
\Dcal_5 &= \mu(\vq_1,\vq_2,\vQ_5)+\mu(\vq_1,\vq_2,\vQ_6)+\mu(\vq_1,\vq_2,\vQ_{10})+\mu(\vq_1,\vQ_5,\vQ_6)+\\
        & \qquad + \mu(\vq_1,\vQ_6,\vQ_{10})+\mu(\vq_2,\vQ_5,\vQ_6)+\mu(\vq_2,\vQ_5,\vQ_{10})+\mu(\vQ_5,\vQ_6,\vQ_{10})\\
\Dcal_6 &= \mu(\vq_1,\vq_2,\vQ_4)+\mu(\vq_1,\vq_2,\vQ_7)+\mu(\vq_1,\vQ_4,\vQ_5)+\mu(\vq_1,\vQ_4,\vQ_7)+ \\
        & \qquad + \mu(\vq_1,\vQ_5,\vQ_7)+\mu(\vq_2,\vQ_4,\vQ_5)+\mu(\vq_2,\vQ_5,\vQ_7)+\mu(\vQ_4,\vQ_5,\vQ_7)\\
\Dcal_7 &= \mu(\vq_1,\vq_2,\vQ_5)+\mu(\vq_1,\vq_2,\vQ_6)+\mu(\vq_1,\vq_2,\vQ_7)+\mu(\vq_1,\vQ_5,\vQ_6)+ \\
        & \qquad + \mu(\vq_1,\vQ_6,\vQ_7)+\mu(\vq_2,\vQ_5,\vQ_6)+\mu(\vq_2,\vQ_5,\vQ_7)+\mu(\vQ_5,\vQ_6,\vQ_7)\\
\Dcal_8 &= \mu(\vq_1,\vq_2,\vQ_5)+\mu(\vq_1,\vq_2,\vQ_6)+\mu(\vq_1,\vq_2,\vQ_{11})+\mu(\vq_1,\vQ_5,\vQ_6)+\\
        & \qquad + \mu(\vq_1,\vQ_6,\vQ_{11})+\mu(\vq_2,\vQ_5,\vQ_6)+\mu(\vq_2,\vQ_5,\vQ_{11})+\mu(\vQ_5,\vQ_6,\vQ_{11}) \ .
\end{aligned}
\end{equation}

\end{document}

%% file: tikz.tex
\newcommand{\arcprop}[6]{
    \pgfmathsetmacro{\firstarrowangle}{ - #3 - (#3 + #4)/2 }   
    \pgfmathsetmacro{\secondarrowangle}{- #4 + (#3 + #4)/2}  
    \draw[thick] (#1) arc (#3:#4:#2) 
        node[currarrow, xscale=-1, sloped, pos=0.25] {} 
        node[currarrow, sloped, pos=0.75] {} 
        node[circ, pos=0.5, label ={[shift={(0,#6)}]#5}] {};
}

\newcommand{\lineprop}[4]{
    \draw[thick] (#1) to
        node[currarrow, xscale=-1, sloped, pos=0.25] {} 
        node[currarrow, sloped, pos=0.75] {} 
        node[circ, pos=0.5, label ={[shift={(0,#4)}]#3}] {} (#2);
}
\tikzset{
  mom/.style={
    postaction={decorate},
    decoration={markings,
      mark=at position #1 with {\arrow{Stealth}}
    }
  }
}

\newcommand{\momArc}[7]{%
  \draw[thick,->] (#1) arc (#2:#3:#4)
    node[sloped, pos=#6, fill=white, inner sep=1pt, yshift=#7] {#5};
}
\newcommand{\momLine}[5]{%
  \draw[thick,->] (#1) -> (#2)
    node[sloped, pos=#4, fill=white, inner sep=1pt, yshift=#5] {#3};
}

\begin{tikzpicture}[]

\def\R{1.75}     
\def\shift{0.3}     
\def\lenratio{1/2}     
\node (L) at (-\R,0) {};
\node (R) at ( \R,0) {};
\node (T) at (0, \R) {};
\node (B) at (0,-\R) {};

\node (Lshifted) at (-\lenratio*\R,\shift) {};
\node (Rshifted) at ( \lenratio*\R,\shift) {};

\arcprop{T}{\R}{90}{0}{}{}
\arcprop{L}{\R}{180}{90}{}{}
\arcprop{B}{\R}{-90}{0}{}{}
\arcprop{L}{\R}{-180}{-90}{}{}

\momArc{0.6,\R+0.2}{80}{15}{\R}{$\vk_4-\vq_1-\vq_2$}{0.55}{10pt}
\momArc{0.6,-\R-0.2}{-80}{-15}{\R}{$\vq_1$}{0.55}{-10pt}
\momArc{-0.6,\R+0.2}{100}{165}{\R}{$\vk_2+\vk_3+\vq_1+\vq_2$}{0.55}{10pt}
\momArc{-0.6,-\R-0.2}{-100}{-165}{\R}{$-\vk_2-\vq_1$}{0.55}{-10pt}

\lineprop{L}{R}{}{}
\momLine{Lshifted}{Rshifted}{$\vq_2$}{0.55}{7pt}

\tikzset{
    sq/.style={draw=black,fill=white,line width=1pt,rounded corners=0pt,
    minimum size=3mm,inner sep=0pt}
}
\node[sq] at (L) {};
\node[sq] at (R) {};
\node[sq] at (T) {};
\node[sq] at (B) {};

\draw[thick] (L) to node[currarrow, xscale = -1] {} (-1.5*\R, 0);
\draw[thick] (R) to node[currarrow] {} (1.5*\R, 0);
\draw[thick] (T) to node[currarrow, sloped] {} (0,1.5*\R);
\draw[thick] (B) to node[currarrow, sloped] {} (0,-1.5*\R);

\node[] at (0.3,1.5*\R+0.2) {$\vk_1$};
\node[] at (0.3,-1.5*\R+0.2) {$\vk_2$};
\node[] at (-1.5*\R-0.1,-0.3) {$\vk_3$};
\node[] at (1.5*\R+0.2,-0.3) {$\vk_4$};

\end{tikzpicture}